\documentclass{pspum-l}

\usepackage{amsthm}
\usepackage{amsmath}
\usepackage{amssymb}
\usepackage{amsfonts}
\usepackage{color}
\definecolor{niceblue}{rgb}{0,0,0.6}
\usepackage[a4paper,colorlinks,citecolor=niceblue,linkcolor=niceblue,urlcolor=niceblue,pdfpagemode=None]{hyperref}

\usepackage{mathrsfs}
\usepackage{latexsym}
\usepackage{dsfont}
\usepackage{color}
\usepackage[all,cmtip]{xy}

\newcommand{\E}{\text{e}}

\newcommand{\C}{\mathds{C}}

\newcommand{\R}{\mathds{R}}

\newcommand{\be}{\begin{equation}}
\newcommand{\ee}{\end{equation}}
\newcommand{\bes}{\begin{equation*}}
\newcommand{\ees}{\end{equation*}}
\newcommand{\del}{\partial}
\newcommand{\Coder}{\text{Coder}}
\newcommand{\MC}{\operatorname{\mathcal{MC}}}
\newcommand{\HH}{\operatorname{HH}}
\newcommand{\Jac}{\operatorname{Jac}}

\newcommand{\id}{\text{id}}

\newcommand{\End}{\operatorname{End}}

\newtheorem{theorem}{Theorem}
\numberwithin{equation}{section}

\begin{document}

\title{An invitation to algebraic topological string theory}

\author{Nils Carqueville}
\address{Arnold Sommerfeld Center for Theoretical Physics, LMU M\"unchen \& Excellence Cluster Universe}
\email{nils.carqueville@physik.uni-muenchen.de}

\author{Michael M.~Kay}
\address{Arnold Sommerfeld Center for Theoretical Physics, LMU M\"unchen \& Excellence Cluster Universe}
\email{michael.kay@physik.uni-muenchen.de}


\begin{abstract}
The purpose of this note is to provide a short invitation to the universal algebraic approach to topological string theory. In the first section we make an attempt to explain the origin of this approach and how it fits into the bigger picture of full string theory, while in the second half of this note we will introduce the relevant notions in more detail and discuss some of our main results on bulk-deformed open topological string amplitudes. 
\end{abstract}

\maketitle

\section{Introduction}

We start our discussion with the bulk sector. In the low energy limit this amounts to the study of $\mathcal N=1$ supergravity in ten dimensions and its solutions, viewed as vacua of closed string theory. We restrict our attention to such solutions that have a six-dimensional compact factor~$M$. Then the Einstein equations imply that this manifold is a Calabi-Yau space. 

In the framework of closed perturbative string theory these solutions~$M$ are realised as the possible targets of two-dimensional sigma models with $\mathcal N=(2,2)$ superconformal symmetry. More precisely, solutions of classical gravity appear as the low energy limit of the worldsheet description. There is a distinguished set of \textsl{marginal} fields of these $\mathcal N=(2,2)$ sigma models that implement infinitesimal deformations which may transform the metric of the corresponding target Calabi-Yau manifold to another Calabi-Yau space closeby. In general the superconformal field theory (CFT) $\mathcal C$ associated to~$M$ is deformed to another $\mathcal N=(2,2)$ CFT, which may be more ``stringy'' in nature in the sense that it does not need to have a geometric interpretation. The marginal fields can be viewed as defining sections of the cotangent bundle of the moduli space of $\mathcal N=(2,2)$ CFTs containing our initial~$\mathcal C$, namely the space of vacua continuously connected to~$\mathcal C$. 

More generally one can consider all \textsl{chiral primary} fields that deform away from~$\mathcal C$ in the much larger moduli space of $\mathcal N=(2,2)$ supersymmetric field theories. Locally (anti-) chiral primaries define Riemann normal coordinates on this moduli space for a neighbourhood of the CFT we started with. We will denote these coordinates by $t_{i}$ and refer to them as closed moduli. 

The chiral primaries can be divided into left- and right-moving zero modes of either the $(c,c)$ (chiral, chiral) or the $(a,c)$ (antichiral, chiral) sectors. The first sector corresponds to type IIB string theory, while the second to type IIA. From the point of view of the superconformal algebra, the two types are related by an involutive outer automorphism. Thus, for every type IIA theory there should be a corresponding type IIB theory -- this is precisely the mirror symmetry conjecture. Of the chiral primaries that deform the metric of the Calabi-Yau, those in the type IIB theory correspond to complex structure deformations, while those in the type IIA theory govern deformations of the complexified K\"ahler structure. Thus by Yau's theorem the type IIA and type IIB theories together contain complete information about the moduli space of metrics. 

Both the $(c,c)$ and $(a,c)$ fields, together with their operator product expansion, naturally form an algebra. In fact these are Frobenius algebras that define two topological field theories (TFTs) called the B-model and A-model, respectively. More generally one can construct two such TFTs from any $\mathcal N=(2,2)$ CFT by a procedure known as topological twisting where the chiral primary fields appear as the cohomology of a BRST operator. The subsector of full string theory that builds on these TFTs is the one that we are interested in here. It is precisely the sector of all chiral primaries, and from it one can compute quantities of the \textsl{full theory} like the effective superpotential that we discuss below. 

We will now consider either one of these topological twists. The associated algebra of fields~$\phi_{i}$ is encoded in the structure constants $\langle\phi_i(0), \phi_j(1)\phi_k(\infty)\rangle_{\text{bulk}}$ where $\langle\,\,\cdot\,,\,\cdot\,\rangle_{\text{bulk}}$ is the topological metric and  $0, 1, \infty$ is the standard choice of punctures on the genus zero worldsheet. Note that the set of structure constants represents only a single point in our moduli space of $\mathcal N=(2,2)$ field theories. What we would like to describe is a (possibly only infinitesimal) patch around that point. The local coordinates of this patch are the closed moduli $t_i$, which allow us to deform the TFT \textsl{correlators} $\langle\phi_i(0), \phi_j(1)\phi_k(\infty)\rangle_{\text{bulk}}$ to the \textsl{amplitudes}
\be\label{deformed-bulk-correlators}
\Big\langle\phi_i(0), \phi_j(1)\phi_k(\infty) \, \E^{\sum_{l}t_l \int \phi_l^{(2)}} \Big\rangle_{\text{bulk}}
\ee
where $\phi_l^{(2)}$ are the two-form descendants of the chiral primaries $\phi_{l}$. 

Let us expand the exponential in the amplitudes and define higher maps~$\ell_n$ which act on the fields~$\phi_{i}$ as follows:
$$
\Big\langle\phi_{i_{0}}(0), \phi_{i_{1}}(1)\phi_{i_{2}}(\infty) \int \phi_{i_3}^{(2)}\ldots \int \phi_{i_n}^{(2)} \Big\rangle_{\text{bulk}}
= \Big\langle \phi_{i_0}, \ell_n(\phi_{i_1} , \ldots , \phi_{i_n})\Big\rangle_{\text{bulk}} .
$$
The topological metric $\langle\,\,\cdot\,,\,\cdot\,\rangle_{\text{bulk}}$ together with the maps~$\ell_n$ define what is called a Calabi-Yau $L_{\infty}$-structure~\cite{z9206084,z9705241,ks0410291} (see the next section for more details). This type of algebra completely encodes the full structure of classical closed topological string theory: knowing the maps~$\ell_n$ and the topological metric one can compute all amplitudes. 

\medskip

Now we turn to the boundary sector. There are several reasons for introducing open strings, ranging from phenomenological to mathematical. For example, from a phenomenological perspective open strings and branes allow for nonabelian gauge symmetries in four-dimensional effective low energy field theories, and their presence also reduces the amount of spacetime supersymmetry. Another compelling reason for introducing open strings appears if one leaves the classical description of gravity by turning on the string coupling constant. It was in fact shown in~\cite{w9306122} that the string coupling constant can be interpreted as the deformation parameter that quantises the moduli space of Calabi-Yau manifolds attached to a given $M$, hence providing a first step to the quantisation of gravity. In this framework open strings are necessary to describe excited states of the quantum theory~\cite{nw0709.2390}. 

In the following we will restrict to the genus zero case which already has a rich structure that demands deeper understanding. We are mostly interested in boundary conditions that describe BPS branes. In the topologically twisted theory they descend to branes whose only open string states are chiral primaries that have an operator product that is strictly associative. Hence these branes and open strings naturally form the objects and morphisms of a category. Together with sewing relations and the boundary topological metric $\langle\,\,\cdot\,,\,\cdot\,\rangle_{\text{bdry}}$, this data combined with the closed TFT structure defines an open-closed TFT as in~\cite{l0010269,ms0609042}. For notational simplicity we will mostly consider the case of only one brane in this note. 

The open and closed structure at one point of our moduli space of $\mathcal{N}=(2,2)$ theories is encoded in the deformed open string three-point correlator
$$
\Big\langle \psi_{a_0}(p_0), \psi_{a_1}(p_1) \psi_{a_2}(p_2) \,\mathcal P\, \E^{\sum_{i}u_{a_{i}} \int\psi_{a_{i}}^{(1)}} \Big\rangle_{\text{bdry}}
$$
which is the boundary version of the bulk sector expression~\eqref{deformed-bulk-correlators}. Here the $\psi_{a}$ denote the open string chiral primaries inserted at generic points $p_0, p_1, p_2$ on the boundary of the disk, and the $\psi^{(1)}_{a}$ are their descendants implementing deformations of the purely open theory. The $u_a$ are free moduli which are to be viewed as local coordinates of a non-commutative space; setting them to zero we obtain the TFT structure constants. 

Just like in the bulk sector we can expand the exponential and consider individual open string amplitudes which are now given in terms of \textsl{higher products}~$\widetilde r_n$:
$$
\Big\langle\psi_{a_{0}}(p_{0}), \psi_{a_{1}}(p_{1})\psi_{a_{2}}(p_{2}) \int \psi_{a_3}^{(1)}\ldots \int \psi_{a_n}^{(1)} \Big\rangle_{\text{bdry}}
= \Big\langle \psi_{a_0}, \widetilde r_{n}(\psi_{a_1} \otimes \ldots \otimes \psi_{a_n})\Big\rangle_{\text{bdry}} .
$$
The topological metric $\langle\,\,\cdot\,,\,\cdot\,\rangle_{\text{bdry}}$ and the maps $\widetilde r_n$ define a \textsl{Calabi-Yau $A_{\infty}$-algebra} which arises from the symmetries of amplitudes and which completely encodes the full structure of open topological string theory~\cite{hll0402, c0412149} (see the next section for the definition). Similarly, for arbitrarily many branes we are led to a Calabi-Yau $A_{\infty}$-category. In the case of the $B$-model with target~$M$ this category is the bounded derived category of coherent sheaves $\mathds{D}^b(M)$, while the boundary sector of the A-model is the derived Fukaya category $\operatorname{Fuk}(M)$. 

Recall that in the purely closed sector the mirror symmetry conjecture says that there is an involution $\mathcal{I}$ acting on the set of moduli spaces of Calabi-Yau manifolds, exchanging complex structure deformations with complexified K\"ahler structure deformations. In the boundary sector the conjecture extends to homological mirror symmetry which in particular states that $\mathds{D}^b(M)$ and $\operatorname{Fuk}(\mathcal{I}(M))$ are equivalent as Calabi-Yau $A_{\infty}$-categories. Thus we have identified a second important reason to study $A_{\infty}$-algebras. 

So far we have sketched how the structure of the full moduli space of an open-closed topological string theory at genus zero consists of $A_{\infty}$-categories fibred over the commutative moduli space of closed TFTs. In the next section we will give a prescription for how to glue these fibres in a neighbourhood of a chosen point in moduli space. In particular we will arrive at an explicit recursive formula for the \textsl{bulk-deformed open string amplitudes}
$$
\Big\langle\psi_{a_{0}}(p_{0}), \psi_{a_{1}}(p_{1})\psi_{a_{2}}(p_{2}) \int \psi_{a_3}^{(1)}\ldots \int \psi_{a_n}^{(1)} \, \E^{\sum_{i}t_i \int \phi_i^{(2)}} \Big\rangle_{\text{bdry}}
$$
which we will express as $\langle \psi_{a_0}, \widetilde r^t_{n}(\psi_{a_1} \otimes \ldots \otimes \psi_{a_n})\rangle_{\text{bdry}}$ in terms of the higher products~$\widetilde r_{n}^t$ of a curved $A_{\infty}$-algebra. Once these amplitudes are computed one immediately obtains their generating function
\be\label{Weff}
\mathcal W_{\text{eff}}(t,u) = \sum_{n \geqslant 2}\frac{1}{n+1} \Big\langle \psi_{a_0}, \widetilde r^t_{n}(\psi_{a_1} \otimes \ldots \otimes \psi_{a_n})\Big\rangle_{\text{bdry}} u_{a_{0}}u_{a_{1}}\ldots u_{a_{n}}
\ee
which is also the F-term D-brane superpotential of the effective four-dimensional low energy field theory. 

Our construction splits into two parts, the first of which is restricted to the case of B-twisted affine Landau Ginzburg models (with arbitrary potential~$W$), while the second part is completely general. Let us mention some of the reasons why Landau-Ginzburg models are interesting theories to study in this context. Firstly, the simple description of their boundary conditions in terms of matrix factorisations allows both for very explicit calculations and for a direct analysis of the underlying structure, unfettered by unnecessary complications. Secondly, by the CY/LG correspondence there is an equivalence between orbifolded Landau-Ginzburg models with quasi-homogeneous potential~$W$ and B-models whose compact targets are hypersurfaces $\{W=0\}$ in projective space, both in the bulk~\cite{w9301} and boundary sector~\cite{hhp0803.2045}. Finally, Landau-Ginzburg models are important since via RG flow they are expected to be related to full CFTs. Several quantities of interest are invariants under the flow and can thus be studied on the often more accessible Landau-Ginzburg side of this CFT/LG correspondence. Again we stress that the CFTs describable in this way can but need not have a geometric interpretation, thus covering a larger class of possible string vacua. 

\medskip

Finally we comment on the relation between the algebraic approach advocated here and other approaches to topological string theory. These are more geometric in nature, both in the sense that they apply only to sigma models with a certain class of target manifolds~$M$ and that they derive the effective superpotential $\mathcal W_{\text{eff}}$ as a geometric quantity. In fact $\mathcal W_{\text{eff}}$ is treated as a commutative function of marginal fields only, and the focus is on computing this function. There are two related geometric approaches which we briefly discuss in turn. 

The most widely used approach aims to generalise the special geometry of the closed string sector~\cite{bcov9309140} to the open and closed B-model, initiated in~\cite{lmv0207259, w0605162}. While a general proof is lacking, in this approach $\mathcal{W}_{\text{eff}}$ is computed as a linear combination of relative period integrals over 3-cycles ending on D2-branes. The computation of these period integrals involves equations of Picard-Fuchs type, and once these are solved the challenge is to find the initial conditions that will yield the correct linear combination of relative periods. This approach works for general toric targets, and it has been applied successfully to several examples with compact Calabi-Yau manifolds. 

A second approach~\cite{bbg0704.2666, bbs1007.2447} uses the CY/LG correspondence and is restricted to a subclass of compact Calabi-Yau manifolds. It assumes knowledge of an explicit family (parametrised by open moduli) of BPS D2-branes in the associated B-model. This family is then transported to the Landau-Ginzburg side where $\mathcal{W}_{\text{eff}}$ can be computed to first order in the bulk moduli simply by computing TFT three-point correlators. By iterating this procedure one may obtain a complete non-perturbative description of the open string moduli space attached to the initial family of D2-branes. 

The geometric approaches to computing $\mathcal{W}_{\text{eff}}$ are very successful and efficient for the classes of models they are applicable to. This level of efficiency has not yet been achieved in the approach via homotopy algebras, which is however mostly due to the fact that it has received much less attention so far. On the other hand, the algebraic approach is founded on the symmetries of amplitudes and thus universally applies to any topological string theory. As we will see below the $A_{\infty}$-structure encoding the amplitudes is derived directly from a Chern-Simons-esque string field theory. In this sense it is also more conceptual, and effective superpotentials~\eqref{Weff} are obtained as a byproduct.

\section{Algebraic approach to bulk-deformed open topological string theory}

In this section we start with a very short introduction to $A_{\infty}$- and $L_{\infty}$-algebras. Then we discuss B-twisted Landau-Ginzburg models and explain how such algebras describe their boundary and bulk sectors, respectively. Finally, we review the construction of bulk-induced deformations of open topological string theory for such models. In an attempt to hide less significant technical details from this exposition, we shall treat signs, degrees, shifts etc.~rather negligently. For a full account we refer to our paper~\cite{ck1104.5438}. 

\subsection{$\boldsymbol{A_{\infty}}$- and $\boldsymbol{L_{\infty}}$-basics}

Let us start by recalling the necessary background on \textsl{algebras with higher structures}. Most importantly, a \textsl{curved $A_{\infty}$-algebra} is a graded vector space~$A$ together with an operator~$\del$ on the space $T_{A}=\bigoplus_{n\geqslant 0} A^{\otimes n}$ satisfying (i) $\Delta \del = (\del\otimes 1 + 1\otimes\del)  \Delta$ and (ii) $\del^2=0$. Here~$\Delta$ is the comultiplication defined by $\Delta (a_{1}\otimes\ldots\otimes a_{n}) = \sum_{j=0}^n (a_{1}\otimes\ldots\otimes a_{j}) \otimes (a_{j+1}\otimes\ldots\otimes a_{n})$, and property (i) tells us that~$\del$ is a \textsl{coderivation}, i.\,e.~the dual notion of a derivation. One can show that it is completely determined by the family of \textsl{higher products}
$$
r_{n} = \pi_{A}   \del\big|_{A^{\otimes n}}: A^{\otimes n} \longrightarrow A \, , \quad n\geqslant 0 \, ,
$$
where $\pi_{A}$ is the projection $T_{A}\rightarrow A$. Accordingly we will use both $(A,\del)$ and $(A,r_{n})$ to denote $A_{\infty}$-algebras. The maps $r_{n}$ are constrained by quadratic relations coming from property (ii). In terms of $C=r_{0}(1)$, $d=r_{1}$ and $a\cdot b = \pm r_{2}(a\otimes b)$ the constraint $\del^2=0$ in particular implies $d(C)=0$, $d^2(a)=a\cdot C - C \cdot a$ and the product rule $d(a\cdot b) = d(a) \cdot b\pm a\cdot d(b)$. In the special case when $r_{n}=0$ for all $n\geqslant 3$ the only additional constraint is that the product $r_{2}$ is associative, and hence the data $(A,C,d,\cdot)$ define a \textsl{curved differential graded (DG) algebra}. 

From the above we see that for any curved $A_{\infty}$-algebra~$r_{1}$ is a differential (to be interpreted as the BRST operator in topological string theory) whenever the \textsl{curvature}~$r_{0}$ is central or vanishes. We call the $A_{\infty}$-algebra \textsl{minimal} if the differential vanishes too. Furthermore, $(A,r_{n})$ is \textsl{cyclic} with respect to a pairing $\langle\,\cdot\,,\,\cdot\,\rangle$ on~$A$ if $\langle a_{0}, r_{n}(a_{1}\otimes\ldots\otimes a_{n})\rangle = \pm\langle a_{n}, r_{n}(a_{2}\otimes\ldots\otimes a_{n}\otimes a_{0})\rangle$ for all $n\geqslant 0$. Finally, an $A_{\infty}$-algebra is \textsl{Calabi-Yau} if it is minimal, cyclic with respect to a non-degenerate pairing, and has a unit compatible with the higher products. As recalled in the previous section, every open topological string theory is described by a Calabi-Yau $A_{\infty}$-algebra. 

We also need the notion of a \textsl{morphism} between curved $A_{\infty}$-algebras $(A,\del)$ and $(A',\del')$. This is a map~$F$ from $T_{A}$ to $T_{A'}$ such that $\Delta'  F=(F\otimes F)  \Delta$ and $F  \del = \del'  F$. Again, $F$ is determined by maps $F_{n}=\pi_{A'}  F|_{A^{\otimes n}}$ subject to certain compatibility conditions with the higher products coming from $F  \del = \del'  F$. We call~$F$ an \textsl{$A_{\infty}$-isomorphism} if~$F_{1}$ is an isomorphism $A\rightarrow A'$. In the case of (non-curved) $A_{\infty}$-algebras~$F$ is called an \textsl{$A_{\infty}$-quasi-isomorphism} if $F_{1}$ induces an isomorphism between the cohomologies of the differentials $r_{1}$ and $r'_{1}$. 

The most fundamental result on (non-curved) $A_{\infty}$-algebras is the \textsl{minimal model theorem}~\cite{k0504437, m9809}. It states that any $A_{\infty}$-algebra $(A,r_{n})$ is $A_{\infty}$-quasi-isomorphic to a minimal $A_{\infty}$-algebra $(H=H_{r_{1}}(A), \widetilde r_{n})$ which is unique up to $A_{\infty}$-isomorphism. 

To see this explicitly, let us restrict to the case where $(A,r_{n})$ is a DG algebra and decompose $A \cong H\oplus B\oplus L$ where $B=\text{Im}(r_1)$ and $L$ is the complement of $\text{Ker}(r_{1})$. Such a decomposition provides us with a map $G=(r_1|_L)^{-1} \pi_B:A\rightarrow A$ which together with $\lambda_{2}=r_{2}$ allows us to recursively define for $n\geqslant 3$: 
$$
\lambda_n = -r_2 (G\otimes 1) (\lambda_{n-1}\otimes 1) -r_2 (1 \otimes G) (1 \otimes\lambda_{n-1}) - \sum_{\genfrac{}{}{0pt}{}{i,j\geqslant 2,}{i+j=n}} r_2 (G\otimes G) (\lambda_i\otimes\lambda_j) \, .
$$
Then the higher products on~$H$ are given by $\widetilde r_{n}=\pi_{H}\lambda_{n}$, and the components of the $A_{\infty}$-quasi-isomorphism $F:(H,\widetilde r_{n})\rightarrow (A,r_{n})$ are the inclusion $F_1:H\hookrightarrow A$ and $F_n=G\lambda_n F_1$ for $n\geqslant 2$. Note that these formulas precisely arise from the Feynman diagrams computed in the topological string field theory $(A,r_{n})$ where~$G$ plays the role of the propagator~\cite{l0107162}. 

\medskip

Just like $A_{\infty}$-algebras are generalisations of associative algebras~$A$ with higher products $r_{n}: A^{\otimes n}\rightarrow A$, $L_{\infty}$-algebras are generalisations of Lie algebras~$V$ with higher brackets $\ell_{n}:V^{\wedge n}\rightarrow V$. Similarly, there are appropriate notions of $L_{\infty}$-morphisms which again can be presented as maps $L_{n}:V^{\wedge n}\rightarrow V'$, subject to certain compatibility conditions. The detailed definitions will not matter to us as the only $L_{\infty}$-algebras relevant in this note are \textsl{DG Lie algebras}, i.\,e.~vector spaces endowed with a graded anti-symmetric bracket that satisfies the super Jacobi identity, and a differential compatible with the bracket. It is however crucial to view them as special $L_{\infty}$-algebras since $L_{\infty}$-morphisms (with higher components) between them will be important in what follows. The root of this fact lies in the relation between deformations and solutions to Maurer-Cartan equations that we discuss next. 

Our aim is to study bulk-deformed open topological string theory. Hence we must explain what a \textsl{deformation} of an $A_{\infty}$-algebra $(A,\del)$ is. By definition it is a map $\delta\in\End(T_{A})$ such that $(A,\del+\delta)$ is a curved $A_{\infty}$-algebra. This means that~$\delta$ is a coderivation, $\delta\in\Coder(T_{A})$, and $(\del+\delta)^2$ must vanish. Thus if we write $[\,\cdot\,,\,\cdot\,]$ for the graded commutator and use $\del^2=0$, we see that~$\delta$ must solve
\be\label{MC}
[\del,\delta] + \frac{1}{2}\, [\delta,\delta] = 0 \, .
\ee

Let us recall that for any DG Lie algebra $(V,d,[\,\cdot\,,\,\cdot\,])$ its associated \textsl{Maurer-Cartan equation} reads $d(\delta)+\frac{1}{2}[\delta,\delta]=0$. We denote its space of formal power series solutions modulo gauge transformations $\delta\mapsto \delta+d(\delta)+[\varphi,\delta]$ as $\MC(V,d,[\,\cdot\,,\,\cdot\,])$. It is an important result~\cite{k9709040} that given an $L_{\infty}$-morphism $L: (V,d,[\,\cdot\,,\,\cdot\,]) \rightarrow (V',d',[\,\cdot\,,\,\cdot\,]')$, the map
\be\label{MCtrans}
\delta \longmapsto \sum_{n\geqslant 1} \frac{1}{n!} \, L_{n}(\delta^{\wedge n}) 
\ee
induces a map $\MC(V,d,[\,\cdot\,,\,\cdot\,])\rightarrow\MC(V',d',[\,\cdot\,,\,\cdot\,]')$. Furthermore, this is an isomorphism if~$L$ is an $L_{\infty}$-quasi-isomorphism. 

We now make the obvious yet crucial observation that the deformation condition~\eqref{MC} is precisely the Maurer-Cartan equation of the DG Lie algebra $\text{Coder}({T_{A}})$ with differential $[\del,\,\cdot\,]$ and bracket $[\,\cdot\,,\,\cdot\,]$. In addition, solving~\eqref{MC} to first order (up to gauge transformations) is identical to computing \textsl{Hochschild cohomology} $\HH^\bullet(A,\del) = H_{[\del,\,\cdot\,]}(\text{Coder}({T_{A}}))$. More importantly, the above result on transporting Maurer-Cartan solutions will allow us to construct all $A_{\infty}$-deformations if we can find a quasi-isomorphic DG Lie algebra whose Maurer-Cartan solutions are known. In the remainder of this note we will review how to do this for the case of Landau-Ginzburg models, where the known Maurer-Cartan solutions will be precisely the space of bulk fields. 

\subsection{Landau-Ginzburg models}

We shall consider the topological B-twist of Landau-Ginzburg models with affine target $X=\C^N$ and potential $W\in R=\C[x_{1},\ldots,x_{N}]$. The \textsl{on-shell} space of states in the bulk sector of such two-dimensional topological field theories is the Jacobian $\Jac(W) = R/(\del_{1}W,\ldots,\del_{N}W)$~\cite{v1991}. This is obtained from the \textsl{off-shell} bulk space of polyvector fields $T_{\text{poly}}=\Gamma(X, \bigwedge T^{(1,0)}X)$ by taking cohomology with respect to the BRST operator $[-W, \,\cdot\,]_{\text{SN}}$. Here we denote by $[\,\cdot\,,\,\cdot\,]_{\text{SN}}$ the Schouten-Nijenhuis bracket, the extension of the Lie bracket to polyvector fields. Note that $T_{\text{poly}}$ together with the BRST differential and the bracket has the structure of a DG Lie algebra. A direct computation shows that its Maurer-Cartan solutions are precisely the on-shell bulk space $\Jac(W)$. 

The boundary sector is defined by matrix factorisations of~$W$, i.\,e.~odd supermatrices~$D$ with polynomial entries such that~$D$ squares to~$W\cdot\id$~\cite{kl0210, bhls0305, l0312}. For simplicity we will only consider one such boundary condition. Then the \textsl{off-shell} space~$A$ is simply the space of all polynomial matrices of the same size as~$D$. The boundary BRST operator is given by the graded commutator $[D,\,\cdot\,]$, and together with matrix multiplication this makes~$A$ a DG algebra $(A,\del)$. By definition its cohomology is the \textsl{on-shell} space~$H$. 

Since the off-shell algebra $(A,\del)$ is an $A_{\infty}$-algebra (describing open topological string field theory for Landau-Ginzburg models), we can apply the minimal model theorem to produce an $A_{\infty}$-structure ~$\widetilde\del$ on~$H$ together with an $A_{\infty}$-quasi-isomorphism $F:(H,\widetilde\del)\rightarrow(A,\del)$. For generic choices of propagators $(H,\widetilde\del)$ will not be Calabi-Yau, but this can be corrected using methods of non-commutative geometry as explained in~\cite{c0904.0862}. Hence we can assume that for any Landau-Ginzburg model and all their branes we can construct a Calabi-Yau $A_{\infty}$-algebra $(H,\widetilde\del)$ that encodes the full structure of open topological string theory. 

\subsection{Bulk-deformed Landau-Ginzburg models}

Now we discuss bulk deformations of $(H,\widetilde\del)$. We saw that any deformation must solve the Maurer-Cartan equation of $(\Coder({T_{H}}), [\widetilde{\partial}, \,\cdot\,], [\,\cdot\,,\,\cdot\,])$, but finding such solutions directly is hard. However, we are only interested in \textsl{bulk-induced deformations} which we define as the image of on-shell bulk fields under an $L_{\infty}$-morphism
\be\label{Lmorph} 
(T_{\text{poly}}, [-W, \,\cdot\,]_{\text{SN}}, [\,\cdot\,,\,\cdot\,]_{\text{SN}}) \longrightarrow (\mathrm{Coder}({T_{H}}), [\widetilde{\partial}, \,\cdot\,], [\,\cdot\,,\,\cdot\,]) \, .
\ee
The left-hand side is the off-shell bulk algebra, and we already know that its Maurer-Cartan solutions precisely form the on-shell bulk space $\Jac(W)$. Hence our remaining task is to construct the $L_{\infty}$-map~\eqref{Lmorph}. We shall do this in two main steps which are Theorems~\ref{Th1} and~\ref{Th2} below. 

To set the stage, let us write the $A_{\infty}$-structure of the off-shell open string algebra $(A,\del)$ as $\del=\del_{1}+\del_{2}$ to emphasise that it is a DG algebra with differential $r_{1}=[D,\,\cdot\,]$ and matrix multiplication~$r_{2}$. There is another natural curved $A_{\infty}$-structure on~$A$ which we write as $(A,\del_{0}+\del_{2})$, where the coderivation~$\del_{0}$ corresponds to the curvature $r_{0}=-W\cdot\id$. Similarly, the polynomial ring~$R$ is also a curved algebra $(R,\widehat\del_{0}+\widehat\del_{2})$, where $\widehat\del_{0}$ corresponds to multiplication with $-W$ and $\widehat\del_{2}$ to the usual product. Now we can take the first step: 
\begin{theorem}[\cite{ck1104.5438}]\label{Th1}
\textsl{There is a sequence of explicit $L_{\infty}$-quasi-isomorphisms}
\begin{align*}
(T_{\mathrm{poly}}, \, [-W, \,\cdot\,]_{\operatorname{SN}}, \, [\,\cdot\,,\,\cdot\,]_{\operatorname{SN}}) 
&  \xrightarrow{\quad\varphi_{1}\quad} (\mathrm{Coder}({T_R}), \, [\widehat{\partial}_0 + \widehat{\partial}_2,\,\cdot\,], \, [\,\cdot\,,\,\cdot\,]) \\
&  \xrightarrow{\quad\varphi_{2}\quad} (\mathrm{Coder}({T_{A}}), \, [\partial_0 + \partial_2,\,\cdot\,], \, [\,\cdot\,,\,\cdot\,]) \\
&  \xrightarrow{\quad\varphi_{3}\quad} (\mathrm{Coder}({T_{A}}), \, [\partial_1 + \partial_2,\,\cdot\,], \, [\,\cdot\,,\,\cdot\,]) \, .
\end{align*}
\end{theorem}

The map $\varphi_{3}$ is simply the adjoint action of the $A_{\infty}$-isomorphism~$T$ (which ``cancels'' the ``tadpole''~$r_{0}$) whose non-vanishing components are $T_{0}=D$ and $T_{1}=1_{A}$, and $\varphi_{2}$ is Morita equivalence for curved DG-algebras. 

It turns out that the map $\varphi_{1}$ deserves more attention; it is an interesting variant of deformation quantisation. We recall that the latter is a method to quantise the algebra of classical observables $C^{\infty}(M, \R)$ on a phase space~$M$, which we take to be $\R^d$. The idea is to deform the commutative, associative multiplication $\widehat r_{2}$ on $C^{\infty}(M, \R)$ to an associative but non-commutative $\star$-product of quantum observables. This deformation again leads to a Maurer-Cartan equation. Kontsevich's solution~\cite{k9709040} was to explicitly construct an $L_{\infty}$-quasi-isomorphism $K : (\Gamma(M, \bigwedge TM), [\,\cdot\,,\,\cdot\,]_{\text{SN}}) \rightarrow (\mathrm{Coder}(T_{C^{\infty}(M, \R)}), [\widehat\partial_2,\,\cdot\,], [\,\cdot\,,\,\cdot\,])$, thus providing a one-to-one correspondence between Poisson structures on ~$M$ and perturbative \mbox{$\star$-products} of quantum observables. 

Note that the map~$K$ is precisely the special case of our map~$\varphi_{1}$ if $W=0$. Indeed, we can show that~$K$ continues to be an $L_{\infty}$-map after ``turning on''~$W$. Furthermore, it also remains a quasi-isomorphism as a direct computation yields $\HH^{\bullet}(R, \widehat{\partial}_0 + \widehat{\partial}_2) \cong \Jac(W)$. Hence we arrive at a generalisation of deformation quantisation, namely that the DG Lie algebra of polyvector fields with differential~$[-W,\,\cdot\,]$ governs deformations of $(R,\widehat\del_{0}+\widehat\del_{2})$. 

\medskip

With Theorem~\ref{Th1} we have explicitly classified all deformations of the off-shell open string algebra $(A,\del)$, and we found that all such deformations are bulk-induced. We stress that this off-shell, string field theoretic result is important by itself as it contains information on descendants of ground states. However, one is also interested in computing amplitudes and effective superpotentials in on-shell open topological string theory. Accordingly, we will discuss how to transport deformations of $(A,\del)$ to those of $(H,\widetilde\del)$. Again, this will be achieved by an $L_{\infty}$-morphism that maps off-shell deformations on-shell via~\eqref{MCtrans}:

\begin{theorem}[\cite{ck1104.5438}]\label{Th2}
\textsl{There is an explicit $L_{\infty}$-map}
$$
(\mathrm{Coder}({T_{A}}), \, [\partial,\,\cdot\,], \, [\,\cdot\,,\,\cdot\,]) \longrightarrow (\mathrm{Coder}({T_{H}}), \, [\widetilde\partial,\,\cdot\,], \, [\,\cdot\,,\,\cdot\,]) \, .
$$
\end{theorem}

The main tool that allows us to construct this map is an $L_{\infty}$-version of the \textsl{homological perturbation lemma}, which is another structure transfer result. Given two complexes $(C_{1},d_{1})$, $(C_{2},d_{2})$ with morphisms $i : (C_2, d_2) \rightarrow (C_1, d_1)$, $p: (C_1, d_1) \rightarrow (C_2,d_2)$ and $h \in \mathrm{End}(C_1)$, we call these data a deformation retraction if $pi=1_{C_{2}}$ and $1_{C_{1}}-ip=d_{1}h+hd_{1}$. In this situation the perturbation lemma states that 
\be\label{HPL}
\delta \longmapsto \sum_{n\geqslant 1} p (\delta h)^n \delta i
\ee
maps a deformation~$\delta$ of $(C_{1},d_{1})$ to a deformation of $(C_{2},d_{2})$, and this can be understood in terms of an $L_{\infty}$-quasi-isomorphism between the endomorphism algebras of the two complexes. 

We now use this method in our setting. It was shown in~\cite{ck1104.5438} that for any $A_{\infty}$-algebra $(A,\del)$ with minimal model $(H,\widetilde\del)$ there is a deformation retraction 
$$
\xymatrix{%
({T_{H}}, \widetilde{\partial}) \ar@<+.5ex>[rr]^-{F}  && ({T_{A}}, \partial) \ar@<+.5ex>[ll]^-{\bar F} \\
}%
\!\!\!\xymatrix{%
{}\ar@(ur,dr)[]^{U} 
}%
\, .
$$
Here $F$ is our usual minimal model quasi-isomorphism, and~$\bar F, U$ are defined recursively by
\begin{align*}
U_n & = -\frac{1}{2}G r_{2} \Big(\sum_{i = 1}^{n-1}(U_i \otimes ( 1_{T_{A}}  + F\bar{F})_{n-i} + ( 1_{T_{A}}  + F\bar{F})_{n-i}\otimes U_i \Big) \, , \\
\bar{F}_n & = -\frac{1}{2}\pi_H r_{2} \Big(\sum_{i = 1}^{n-1}(U_i \otimes ( 1_{T_{A}}  + F\bar{F})_{n-i} + ( 1_{T_{A}}  + F\bar{F})_{n-i}\otimes U_i\Big) \, .
\end{align*}
Furthermore, these maps are such that~\eqref{HPL} maps coderivations to coderivations. 

\medskip

Now we have everything in place to apply the above to the case where $(A,\del)$ and $(H,\widetilde\del)$ are the off-shell and on-shell open string algebras. It follows from Theorem~\ref{Th1} that all deformations~$\delta$ of $(A,\del)$ are coderivations determined by $1\mapsto \sum_{i} t_{i}\phi_{i}$ where~$\phi_{i}$ are on-shell bulk fields in $\Jac(W)$ and~$t_{i}$ are the closed moduli. Then Theorem~\ref{Th2} tells us that the associated deformations~$\widetilde\delta$ of the on-shell algebra $(H,\widetilde\del)$ are given by
$$
\widetilde{\delta} = \sum_{n \geqslant 1}\bar{F} (\delta U)^n \delta F \, .
$$
All the ingredients $F,\bar F,U$ can be algorithmically computed, thus leading to completely explicit expressions for the bulk-deformed open string amplitudes
\begin{align*}
\Big\langle \psi_{a_0}, \widetilde r^t_{n}& (\psi_{a_1} \otimes \ldots \otimes \psi_{a_n})\Big\rangle_{\text{bdry}}  \\
& =\Big\langle\psi_{a_{0}}(p_{0}), \psi_{a_{1}}(p_{1})\psi_{a_{2}}(p_{2}) \int \psi_{a_3}^{(1)}\ldots \int \psi_{a_n}^{(1)} \, \E^{\sum_{i}t_i \int \phi_i^{(2)}} \Big\rangle_{\text{bdry}}
\end{align*}
where the higher products $\widetilde r_{n}^t$ are the components of $\widetilde \partial + \widetilde \delta$. 

\subsection*{Acknowledgements} 
We thank I.~Brunner, D.~Murfet and A.~Recknagel for helpful comments on the manuscript.

\end{document}